\documentclass{optica-article}

\journal{opticajournal} 

\articletype{Research Article}

\usepackage[normalem]{ulem} 

\usepackage{lineno}
\usepackage{float}

\begin{document}

\title{Thermal Lensing Effects in Two-Photon Light-Sheet Microscopy}

\author{Antoine Hubert\authormark{1,2}, Hugo Trentesaux\authormark{1,2}, Thomas Pujol\authormark{1,2 }, Georges Debrégeas\authormark{1,2}  and Volker Bormuth\authormark{1,2,*}}

\address{\authormark{1}Sorbonne Université, CNRS, Laboratoire Jean Perrin, LJP, F-75005 Paris, France\\
\authormark{2}Sorbonne Université, CNRS, Inserm, Institut de Biologie Paris-Seine, IBPS, F75005 Paris, France}

\email{\authormark{*}volker.bormuth@sorbonne-universite.fr} 



\begin{abstract*}
In light-sheet fluorescence microscopy (LSFM),  the axial resolution is governed by the illumination beam profile, motivating the development of advanced beam-shaping techniques to enhance imaging performance. Two-photon LSFM (2P-LSFM), in particular, improves the signal-to-background ratio by reducing laser scattering and distortion in biological specimens. However, we report a potentially detrimental thermal effect in 2P-LSFM: the high laser powers required for two-photon excitation induce localized heating, which alters the refractive index of the medium and effectively forms a divergent thermal lens in water. At 500 mW the light-sheet waist broadens by 25\% and shifts by 300 µm before stabilizing several seconds after the laser shutter is opened. Both experiments and simulations reveal that this thermal lensing effect scales with laser power and the path length the beam travels through water. The resulting degradation in resolution and signal-to-noise ratio may compromise imaging applications that require high laser powers for rapid volumetric imaging of large specimens or functional brain imaging. This limitation is particularly critical in dynamic sample environments, such as during stepwise repositioning or flow-based delivery of chemical or hydrodynamic sensory stimuli, where changes occur on timescales comparable to the thermal settling time.
\end{abstract*}



\section*{Introduction}
Since its introduction more than two decades ago \cite{huisken2004optical}, light-sheet fluorescence microscopy (LSFM) has become a key imaging technique in developmental biology and neuroscience, enabling high-resolution imaging of \textit{in vivo} and cleared samples \cite{keller2008reconstruction, voigt2019mesospim, yang2022daxi}. Its ability to perform rapid volumetric acquisition makes it well-suited for functional imaging combined with behavioral studies, or large field-of-view imaging at single-cell resolution in 3D \cite{keller2008reconstruction, panier2013fast, ahrens2013whole, migault2018whole}.

Multiphoton light-sheet microscopy (2P-LSFM or 3P-LSFM) has later emerged as a powerful alternative for denser tissue by offering reduced scattering and greater penetration depth due to infrared excitation \cite{truong_deep_2011,fahrbach_light-sheet_2013, mahou2014multicolor, escobet2018three}. However, like all multiphoton techniques, it imposes constraints related to photodamage and heat deposition \cite{maioli_fast_2020}. 
Beyond these biological limitations, beam propagation through aqueous media, particularly in long working distance configurations, can introduce additional thermal effects that pose questions about their influence on imaging performance.

Thermal effects are documented in various applications involving high-power lasers, typically in the range of tens to hundreds of watts. For instance, acousto-optic modulators (AOMs) crystals made of $\text{TeO}_2$ are known to induce a converging lens effect within the material, resulting in focal mismatches \cite{simonelli2019realization}. In applications such as optical trapping and laser cutting or optical cavities, these thermal effects must be carefully considered to ensure precise beam control and system performance \cite{loiko2014thermo, simonelli2019realization}. The effects of laser-induced thermal lenses on liquids can also be used as a means to measure the thermo-optical coefficient of a solution \cite{pilla2009measurement}.

In this study, we aim to provide a detailed experimental characterization of the thermal lens effect in the specific context of 2P-LSFM. We examined the relationship between waist size and focal position in relation to laser power and the propagation distance of the laser through water. Additionally, we characterized the settling dynamics and stability of the thermally induced focal shift from the time the laser was switched on. We then developed and implemented simulations to investigate these dynamics further. Integrating experimental observations with simulations allowed us to identify key factors affecting the stability and shape of high-power laser beams focused through liquids, highlighting critical considerations for optimizing performance in 2P-LSFM applications.

\section*{Material and methods}

To investigate the impact of thermal lensing on beam propagation in 2P-LSFM, we directly visualized the laser forming the digitally scanned light sheet using the microscope’s detection objective (see Figure \ref{Figure_Experiment}A). The laser beam was propagated through a water-filled tank containing a low concentration of saturated Rhodamine 6G, which enabled two-photon excitation of the dye. This configuration allowed us to image the fluorescence---and hence the laser beam---perpendicular to its propagation axis.

Our infrared source was a pulsed infrared laser (Coherent, Monaco, 40W, 1035 nm, 273 fs pulses, 50 MHz repetition rate). The light sheet was formed using a 5× IR objective (Olympus LMPLN5XIR, NA = 0.1, WD = 16 mm) that focused the laser first through air and then through a thin cover glass (130-160 µm thickness, Menzel Gläser \#1, Thermo Fisher) into the water tank. The water tank was mounted on a translation stage (Newport, M-UMR12.40) to adjust the laser path length through water. Fluorescence was collected with a 20X water-immersion objective (Olympus XLUMPLFLN20x, NA = 1.0, WD = 2 mm) mounted on a Z-piezo stage (Jena, PZ 400 SG OEM). The signal passed through a short-pass filter (Semrock, FF01-750/SP-25) to block infrared light and a fluorescence emission filter (Semrock, FF01-630/69-25). The image was formed by a tube lens (Thorlabs, TTL180-A) and captured by an sCMOS camera (Hamamatsu C14440-20UP, 6.5 µm/pixel), yielding a field of view of 665 × 748 µm$^2$. We restricted imaging to a region of interest (ROI) around the laser, measuring 93 × 748 µm$^2$. The detection arm (comprising the imaging objective, tube lens, and camera) was also mounted on a translation stage to center the waist of the imaged beam within the field of view.

The light-sheet microscope system was controlled using a custom Matlab interface (R2020A, MathWorks) with a DAQ card (NI USB-6259, National Instruments) for data acquisition.

Beam propagation and waist dynamics were recorded under varying laser powers. 

For the analysis of the steady-state experiments, we acquired an image stack of the laser beam consisting of 40 layers with 1 $\mu$m axial increments. We computed the axial projection of this stack and, for each line, extracted the position of maximum intensity. The beam waist was extracted as the distance from this maximum to the position where the intensity dropped to 1/e$^2$ of the peak value.

For the analysis of the dynamic experiments, we recorded a single plane to achieve high temporal resolution. The waist position was extracted as the location of maximum fluorescence intensity.


\section*{Experimental Results}

\subsection*{Steady-State Thermal Lensing: Focal Shift and Waist Broadening}

\begin{figure*}
\centering
    \includegraphics[scale=0.45]{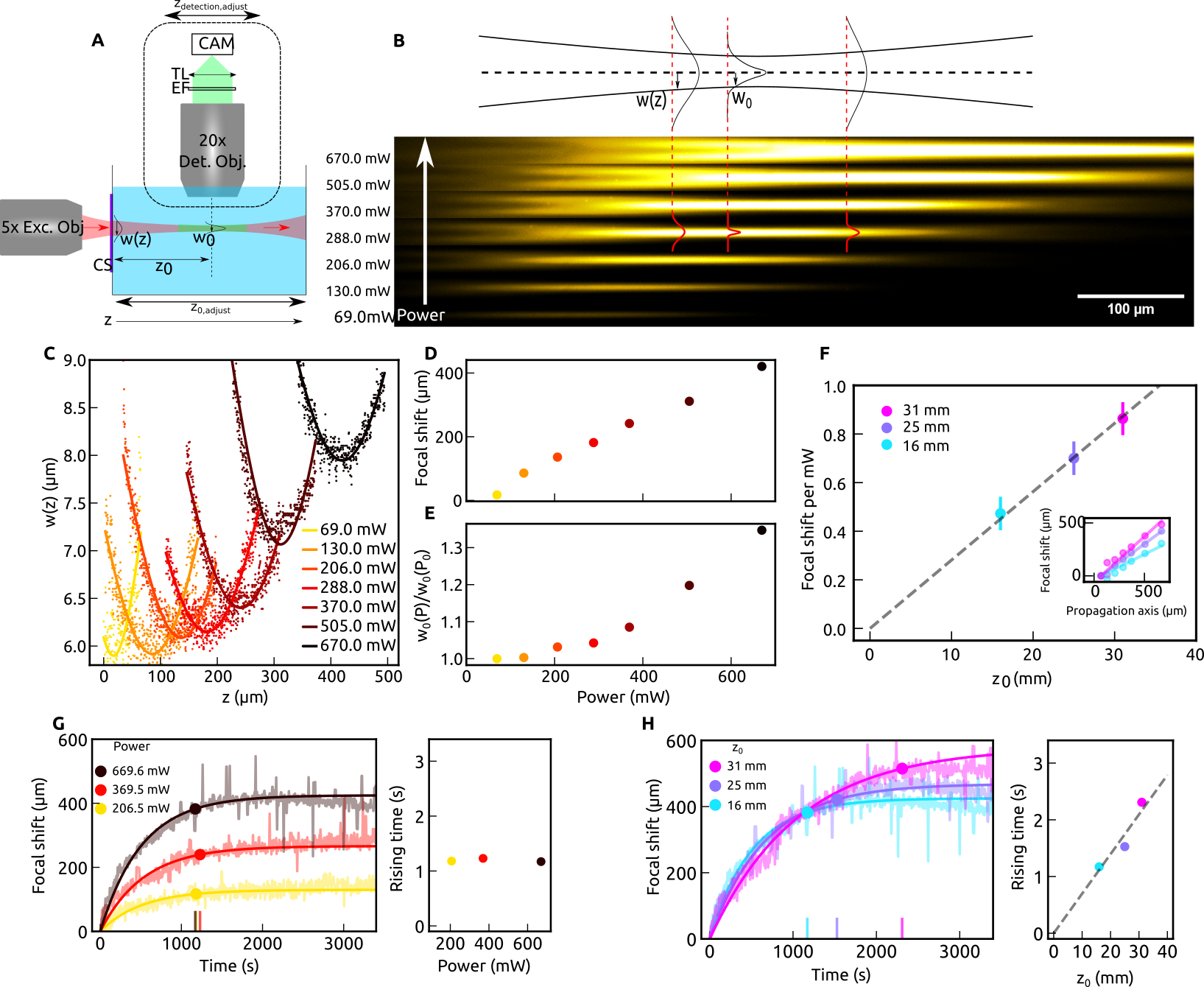}
    \caption{
    (A) Profile and characteristic of the laser beam with increasing power, $z_0$ = 16mm, RR = 50MHz and $\lambda$ = 1035nm. 
    (B) Images vertically stacked of beam propagating through water with increasing power. 
    (C) Beam radius for several powers computed at $1/e^2$ from raw data and associated polynomial fit.
    (D) Focal shift in function of power from the previous fit 
    (E) Waist in function of power from fits of (C) (F) Evolution of the focal shift with power for several $z_0$.
    (G) Left: Temporal evolution of the waist position for a laser beam in water, $\lambda$ = 1035 nm, RR = 50 MHz, $z_0$ = 16 mm, framerate = 185 Hz, exponential fit (solid) based on maximum value reached from raw data (transparent) for three different power. Dots represent the 90\% of max value of the exponential fit and associated so-called rising time. Right: Rising time for different power estimated on exponential fit from (left, error bars are calculated as standard error from the fit 
    (H) Left: Temporal evolution of the waist position for a laser beam in water, $\lambda$ = 1035 nm, RR = 50 MHz, Power = 669.6 mW, framerate = 185 Hz, exponential fit based on maximum value reached (solid) for three different $z_0$. Right: Rising time for different $z_0$ estimated on exponential fit from (left), error bars are calculated as standard error from the fit
    }
    \label{Figure_Experiment}
\end{figure*}

Figure \ref{Figure_Experiment}B shows fluorescent images of the laser beam recorded at various laser powers after transient effects had subsided. As the laser power increases, the focal point shifts along the direction of propagation while the beam waist expands. To quantify these changes, we determined the beam diameter w(z)---the radial distance where the laser intensity dropped to 1/e$^2$---as a function of the position along the propagation direction, z. The resulting axial profiles for different laser powers are shown in Figure \ref{Figure_Experiment}C, with the focal position defined as the location of minimum beam width. 

Within the tested power range of 50–700 mW, the focal shift---defined as the difference in focus position relative to the lowest power configuration---increased linearly with laser power, reaching shifts of up to 0.5 mm at 700 mW (Figure \ref{Figure_Experiment}D). A supralinear expansion of the beam waist accompanies this focal shift. We measured a relative enlargement of the beam waist of $\sim 40\%$ at maximum power (Figures \ref{Figure_Experiment}E). In the absence of thermal lensing effects, the expected waist in water is 3.3 $\mu$m, calculated from the far-field beam divergence measured in air without the water tank. Our fluorescence-based measurement approach systematically overestimates the beam waist due to scattering effects (see Supplementary Note \ref{supp_WaistEstimation}), particularly at small values of $w_0$. Thus, while the observed waist expansion is a genuine effect, its relative magnitude may in fact be underestimated due to this measurement bias.

Because thermal lensing can occur along the entire optical pathway, we further hypothesized that if local heating of the water is the underlying cause, the magnitude of the effect should depend on the optical path length through water. To test this, we repeated the experiments with three different path lengths: the original $z_0$ = 16 mm, $z_0$ = 25 mm, and $z_0$ = 31 mm. In all cases, a linear increase in focal shift with laser power was observed, with the slope of this increase becoming linearly steeper as $z_0$ increased (Figure \ref{Figure_Experiment}F).

In summary, at steady state, the focal shift increased linearly with laser power, with a sensitivity proportional to the optical path length through water. In contrast, the waist broadening exhibited a supralinear dependence on power.

\subsection*{Temporal Dynamics of the Focal Shift}

Figure \ref{Figure_Experiment}G illustrates the temporal evolution of the beam waist position immediately after the laser is switched on. We used high-speed imaging to monitor the dynamic change in beam profile as the focal shift gradually approaches its steady state value. We found that the focal shift converges quasi-exponentially with a characteristic time constant on the order of a few seconds. Although the final focal shift increases with laser power, the transient dynamics remains largely independent of power, with a mean time constant of $\tau$ = 1.19 $\pm$ 0.015 s. Notably, increasing the laser propagation length through water ($z_0$) leads to a linear increase in the time constant required to reach steady state (Figure \ref{Figure_Experiment}H).

\section*{Numerical Simulations}

\subsection*{Thermal Lens Model}

\begin{figure*}
\centering
\includegraphics[width = 10 cm]{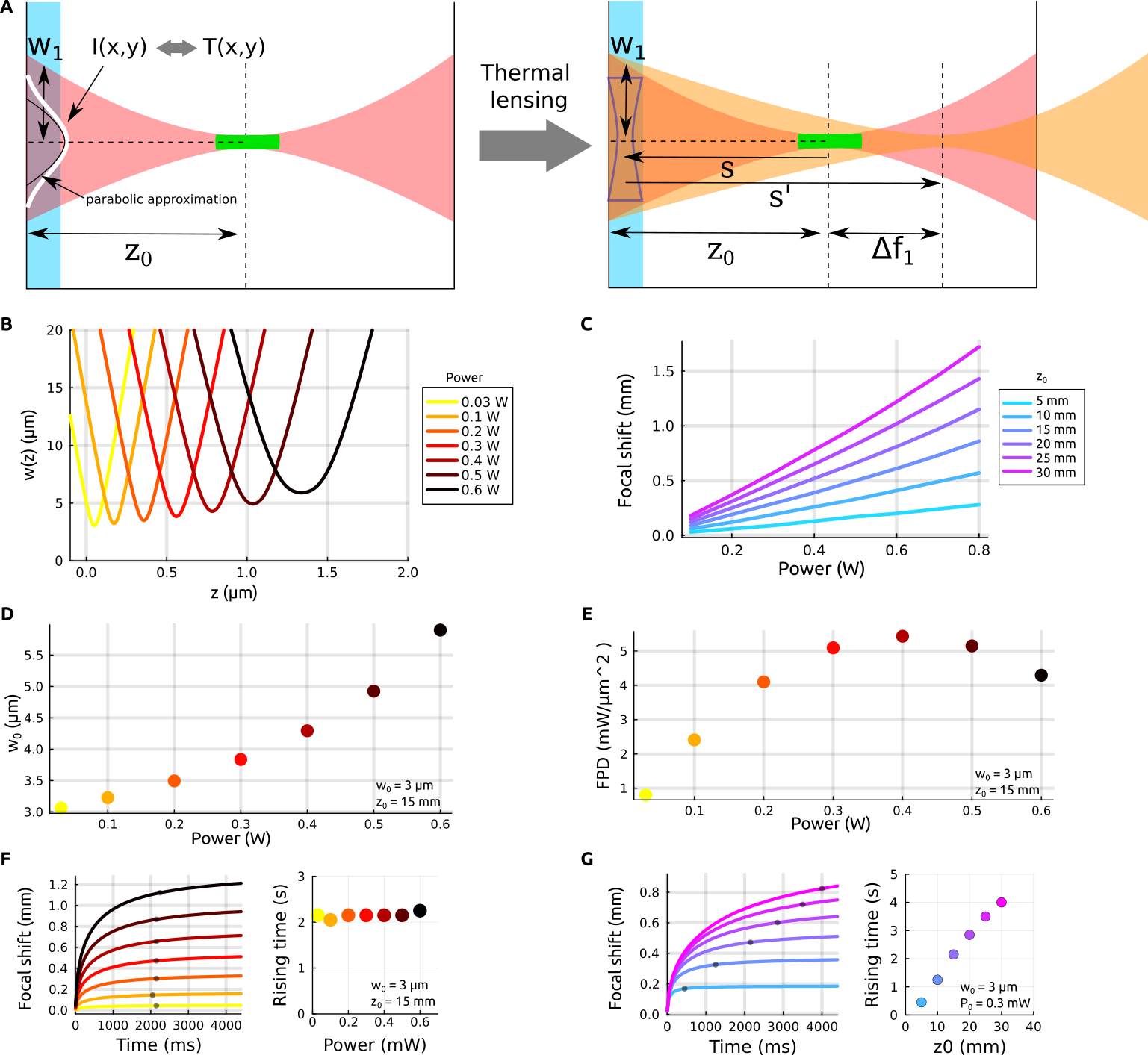}
\caption{Schematic of the model describing thermally induced focal shift.
(A) A radial temperature gradient (white) is induced by the laser beam intensity profile (red). The path length through water is denoted $z_0$. We begin by modeling a thin water layer (blue), with the green area representing the initial focal region used in typical 2P-LSFM. This thin water layer acts as a weak diverging lens, producing a focal shift $\Delta f_1$.
(B) Beam profiles for increasing laser power. 
(C) Focal shift as a function of power for increasing $z_0$. 
(D) Laser waist at focus as a function of laser power.
(E) Focal power density (FPD) as a function of laser power.
(F) Temporal evolution of the focal for different $z_0$.
(G) Temporal evolution of the focal for different laser powers.
The grey dots in F and G represent the time to reach 90\% of the maximum value calculated at $t=3$s
}
\label{Fig_Theory}
\end{figure*}

When a Gaussian laser beam propagates through a medium with a finite absorption coefficient, localized heating occurs due to energy deposition. This generates a paraxial temperature increase $T$, which alters the refractive index $n$ via the thermo-optic effect. In water and most liquids, the thermo-optic coefficient $\mathrm{d}n/\mathrm{d}T$ is negative, meaning that the refractive index decreases with increasing temperature. Consequently, the refractive index is reduced along the beam axis and increases radially, forming an effective diverging lens---a phenomenon illustrated in Fig.~\ref{Fig_Theory}.

Gordon et al. showed that the refractive index profile induced by absorption of a Gaussian laser beam of waist $w$ in a medium of initial refractive index $n_0$ can be approximated near the optical axis as a parabola:

\begin{equation}
n(r) = n_0  + \delta \cdot r^2 
\label{eq:refindex}
\end{equation}

Within the thin-lens approximation, the corresponding focal length of the thermal lens at steady state is given by:

\begin{equation}
F(w) = - \frac{n_0}{l} \frac{1}{2 \delta}= \frac{n_0}{l} \frac{\pi k w^2}{P \alpha l \frac{dn}{dT}}  
\label{eq:ThermalLens}
\end{equation}

Here, $l$ is the thickness of the absorbing element, $k$ its thermal conductivity, $w$ the beam waist at the position of the element, $P$ the laser power, $\alpha$ the absorption coefficient, and $\frac{dn}{dT}$ the thermo-optic coefficient. 

To determine the focal shift introduced by a thin thermal lens located at a distance $s_i$ from the laser waist (with $s_i < 0$ indicating a position before the waist), we apply the standard Gaussian optics lens formula \cite{simonelli2019realization}:

\begin{equation}
\Delta f_i = s_i + s_i’ = s_i + \frac{\frac{z_r^2}{F} - s\left(1 - \frac{s_i}{F} \right)}{\frac{z_r^2}{F^2} + \left(1 - \frac{s_i}{F} \right)^2}
\label{eq_FocalShift}
\end{equation}

where $z_r = \pi w_0^2 / \lambda$ is the Rayleigh range, and $s_i{\prime}$ is the distance from the lens to the new image plane of the beam waist.

Crucially, because the beam is focused inside the water volume, the beam waist varies along the propagation axis. Consequently, each water layer introduces a local, thin diverging lens whose effect depends on the waist size at that location, which itself is modified by preceding thermal lenses. As a result, the overall focal shift arises from these thermal lenses' cumulative, interdependent effects. We numerically computed the total focal shift, \(\Delta f = \sum_i{\Delta f_i}\), from this cumulative effect using an iterative scheme with 10 µm-thick water layers for steady state simulations and 1 µm-thick layers for simulations of temporal dynamics.

Gordon et al. also derived the characteristic time $\tau$ over which the thermal lens forms under continuous laser heating, showing that it is governed by heat diffusion over the characteristic length scale set by the beam waist:

\begin{equation}
\tau = \frac{w^2}{4D}
\end{equation}

where $D = \frac{\kappa}{\rho c_p}$ is the thermal diffusivity of the medium, with $\kappa$ denoting thermal conductivity, $\rho$ the density, and $c_p$ the specific heat capacity.

Accordingly, the focal length of the thermal lens under continuous laser illumination at constant power evolves over time, approaching its steady-state value as:

\begin{equation}
F(t, w) = F_{w} \left[1 + \frac{\tau}{2t} \right]
\label{eq:ThermalLens}
\end{equation}

Using this formula in our iterative simulation code, we also simulated the temporal evolution of both the focal shift and the waist broadening.

For all simulations, we used a beam waist $w_0$ = 3.0 $\mu$m, consistent with the theoretical value expected from our experimental setup described above. All other parameters used in the model are listed in Table~\ref{Tabel_SimulationParameters}.

\subsection*{Simulation Results}

\begin{figure*}
    \centering
\begin{tabular}{ l | l | l}
$\lambda$ & 1035 nm   & wavelength \\
$\kappa$ & 0.6W.kg$^{-1}$.K$^{-1}$  & thermal diffusion coefficient \\
$\alpha$ & 19m$^{-1}$                & absorption coefficient at 1035 nm (from \cite{hale1973optical}) \\
$\frac{dn}{dT}$ & -1.1 $\times$ 10$^{-4}$ \, $\mathrm{K}^{-1}$ & thermo-optic coefficient at 20 °C\cite{abbate1978temperature}\\
$c_p$ & 4186J.kg$^{-1}$.K$^{-1}$    & heat capacity \\
$\rho$ & 1000 kg.m$^{-3}$    & density \\

\end{tabular}
\caption{Simulation Parameters }
\label{Tabel_SimulationParameters}
\end{figure*}

\subsubsection*{Steady-State Thermal Lensing: Focal Shift and Waist Broadening}

Our simulations (see Figure \ref{Fig_Theory} and Table \ref{Tabel_SimulationParameters}) successfully reproduce the scaling of both the axial shift and distortion of the laser intensity profile that give rise to focal shift and waist broadening as laser power increases.

Consistent with experimental observations, the simulated focal shift exhibits a linear dependence on laser power, with a sensitivity that increases linearly with $z_0$, the distance traveled by the laser before reaching its focal point.
This result is, at first glance, counterintuitive. According to Eq. \ref{eq:ThermalLens}, the focal length of the induced thermal lens is proportional to the Gaussian beam's waist. The Gaussian beam has a large waist at large distances from the geometric focal point, resulting in a weakly divergent thermal lens with a long focal length. One might therefore expect the contribution of these distant water layers to the overall focal shift to be negligible. However, for $F \gg s$, $w \gg w_0$, $F \gg z_R$, and $t \gg \tau$, the focal shift given by Eq. \ref{eq_FocalShift} induced by each additional layer becomes approximately constant and is given by:

\begin{equation}
 \Delta f \approx \frac{\alpha}{k \lambda} \frac{dn}{dT} P
\end{equation}

This explains the observed linear increase in the cumulative focal shift with increasing optical path length through water.

The model also captures the supralinear increase in waist broadening with laser power, which we observed experimentally.
The model predicts that, as a direct consequence of this dependency, the power density at the focus (FPD) may reach a maximum because laser power increases linearly while the waist broadens supralinearly (see Figure \ref{Fig_Theory}E). Beyond this maximum point, further increases in laser power should lead to a reduction in fluorescence excitation via the two-photon effect, despite the higher total power.

\subsubsection*{Temporal Dynamics of the Focal Shift}

Our simulations also reproduce the quasi-exponential approach to steady state, with a time constant that is independent of laser power and scales linearly with $z_0$, reflecting the cumulative thermal effect along the optical path.

Since $\tau \propto w^2$, the thermal lens near the focal point---where the beam waist is smallest---reaches steady state within a few milliseconds. In contrast, water layers located several millimeters or centimeters away from the focus, where the beam waist is much larger, require significantly longer timescales (up to seconds) to equilibrate. This implies that the most distant water layers define the time required to reach thermal equilibrium. The parameter $z_0$---the distance traveled through water before reaching the focus---is thus a key determinant of the system’s temporal response.


\section*{Conclusion and Discussion}
This study demonstrates that thermal lensing markedly alters the beam waist in 2P-LSFM, both in its position and size. As laser power increases, we experimentally observe a linear focal shift of up to 0.3 mm at 500 mW--power levels typical for fast volumetric calcium imaging in larval zebrafish--and a nonlinear beam waist broadening, reaching approximately 20\% of its original diameter. These distortions degrade optical sectioning, and can introduce fluorescence fluctuations over time, potentially confounding the interpretation of functional recordings.

Our results further indicate that the extent of thermal lensing is strongly influenced by the laser’s propagation distance through water. Longer water paths exacerbate the lensing effects, which is particularly critical for imaging large samples or when a large field of view is required. This dependency emphasizes the need for careful optical design in systems where extended water paths are unavoidable.

To better understand these effects, we developed a theoretical model that captures the spatial and temporal scaling of thermal lensing as functions of laser power, propagation distance, and time. While this simplified model qualitatively reproduces our experimental observations, it systematically overestimates the magnitude of the focal shift by a factor of \textasciitilde2. This discrepancy arises, in part, from our approximation of the induced temperature profile---and consequently the refractive index gradient---being parabolic rather than gaussian like. This assumption holds primarily within radial distances smaller than the laser beam waist. At larger radii, the actual temperature profile deviates from the ideal parabola, leading to a reduction in effective lensing and the introduction of spherical aberrations. These aberrations distort the axial Gaussian beam profile and thereby attenuate the observed focal shift.

Moreover, our iterative numerical approach assumes a Gaussian input beam profile throughout propagation, even though the beam accumulates spherical aberrations while propagating through the heated water. This simplification may further contribute to the mismatch between theory and experiment. Additional sources of aberration, such as the glass slide at the entrance of the water tank, are also neglected in our model. Finally, at the upper end of tested laser powers, convective flows are likely to become significant. Indeed, we observe signatures of convection in the temporal evolution of the focal shift, which reaches a maximum and then gradually diminishes after approximately two seconds—suggesting dynamic redistribution of the heated fluid.

Increasing the mean excitation power is a common strategy in multiphoton imaging to enhance signal intensity, reduce illumination time, and accelerate volumetric acquisition. However, this approach is fundamentally limited by well-known biological constraints, including phototoxicity, photobleaching, and heat-induced damage to the sample. Our results add an important optical consideration: even at laser powers typically used for physiological imaging, thermal lensing introduces significant beam degradation, which can compromise both image quality and spatial resolution. This effect may be especially problematic in advanced illumination techniques that rely on complex beam shaping, such as Bessel-beam light sheets, where thermally induced aberrations could severely disrupt the optical sectioning performance.

These limitations are directly relevant to the practical workflow of 2P-LSFM. A notable challenge arises during the transition from low-power alignment (\textasciitilde50 mW) to high-power imaging conditions (400–500 mW). At low power, precise centering of the laser waist within the field of view is readily achieved. However, under high-power conditions, thermal lensing induces both lateral displacement and broadening of the waist, leading to resolution loss, particularly in the central imaging region. This effect underscores the importance of accounting for power-dependent optical distortions during system alignment and calibration.

Beyond steady-state effects, our study highlights the dynamic nature of thermal lensing. The beam waist evolves over time, following a quasi-exponential saturation with a characteristic time constant on the order of seconds. These transient dynamics have the potential to interfere with imaging in time-varying sample environments. For example, during stepwise repositioning in large-volumetric scans, thermal lensing could introduce spatial inhomogeneities in the illumination pattern across the sample. This concern becomes even more relevant when considering the extension of 2P-LSFM to specific configurations of functional imaging of neural activity. Several advanced experimental paradigms---such as flow-based delivery of chemical \cite{Candelier.2015} or hydrodynamic stimuli \cite{Vanwalleghem.2020} , whole-microscope rotation for vestibular stimulation \cite{migault2018whole,migault2024distinct} , and setups that allow for functional imaging in freely swimming \cite{Kim.2017qlr}---have been successfully implemented using one-photon light-sheet microscopy, where thermal lensing is negligible due to the lower absorption of visible light. However, adapting these strategies to two-photon configurations would increase sensitivity to heat-induced refractive index changes in the medium. In such contexts, dynamic focal shifts caused by thermal lensing could significantly compromise signal fidelity, particularly when the timescale of stimulus delivery or sample movement approaches the thermal settling time. Under these conditions, fluorescence changes arising from beam waist drift may coincide with stimulus-evoked or behaviorally driven neural activity, confounding the interpretation of calcium signals and introducing artifacts into functional recordings. These considerations emphasize the importance of accounting for thermal dynamics when designing experimental protocols for use with high-power 2P-LSFM systems.

The implications of our findings extend beyond passive mitigation of image degradation. A deeper understanding of the interplay between laser power, propagation distance, and thermal dynamics could pave the way for actively controlling beam propagation in high-power 2P-LSFM systems. For instance, adaptive optics or real-time feedback mechanisms could be used not only to restore image quality by correcting for focal shifts and beam broadening, but also to enable more stable and precise illumination conditions.

In summary, this work is, to our knowledge, the first report of thermal lensing in light-sheet fluorescence microscopy. The phenomenon, emerging under high-power IR excitation and extended water paths, may have been overlooked in traditional high-resolution setups that employ short working distances and high-NA objectives. Ultimately, a thorough understanding and effective control of thermal lensing are essential for optimizing 2P-LSFM performance, particularly in demanding applications such as functional imaging in live specimens. Future work should focus on integrating compensation techniques to mitigate these thermal effects, ensuring reliable and high-quality imaging outcomes.


\section*{Code and data availability}
Analysis scripts, data underlying the figures, and simulation code are available on GitHub:\\
\href{https://github.com/LJPZebra/ThermalLens.git}{https://github.com/LJPZebra/ThermalLens.git}

\section*{Conflict of interest statement}
The authors declare no conflict of interest.\\

\section*{Acknowledgements}
We are grateful to Carounagarane Dore for his contribution to the design of the experimental setup. This project has received funding from the European Research Council (ERC) under the European Union's Horizon 2020 research innovation program, grant agreement number 715980. Furthermore, it was supported by grants from Région Ile-de-France, specifically by DIM cerveau et pensée and DIM-ELICIT. Moreover, the project received partial funding from the CNRS and Sorbonne Université. AH was partially funded by a PostDoc fellowship form the i-Bio initiative at Sorbonne University, GM had a Ph.D. fellowship from the Doctoral School in Physics, Ile de France (EDPIF), and HT had a Ph.D. fellowship from Ile de France (ARDoC 17012950).

\section*{Contributions}
AH, HT, TP, GD, and VB designed the project, conducted research, analyzed data, performed simulations, and wrote the manuscript.

\section*{Bibliography}
\bibliography{Biblio}

\onecolumn
\newpage

\section*{\large Supplemental Materials: Thermal Lensing Effects in Two-Photon Light-Sheet Microscopy}

\setcounter{equation}{0}
\setcounter{figure}{0}
\setcounter{table}{0}
\setcounter{page}{1}
\makeatletter
\renewcommand{\thepage}{S\arabic{page}}
\renewcommand{\theequation}{S\arabic{equation}}
\renewcommand{\thefigure}{S\arabic{figure}}
\setcounter{section}{1}
\renewcommand{\thesection}{S - \arabic{section}}
\renewcommand{\thetable}{S\arabic{table}}

\subsection{Laser waist estimation}
\label{supp_WaistEstimation}

\begin{figure}[H]
\centering
\includegraphics[width=0.75\linewidth]{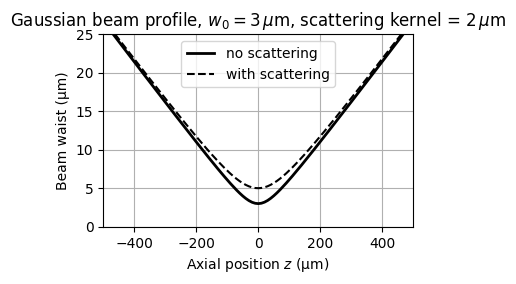}
\caption{Effect of scattering on beam waist estimation.}
\label{fig_ScatteringEffect}
\end{figure}

We overestimate the waist relative to the expected value given the optical properties of our setup. One possible explanation is scattering of fluorescence photons as they travel through the rhodamine solution before reaching the camera, particularly if the dye is not fully dissolved and forms aggregates or micelles. Simulations indicate that even modest scattering of photons emitted from the same plane can substantially broaden the measured beam profile, leading to an overestimation of the beam waist, $w_0$ (see Figure \ref{fig_ScatteringEffect}). In our analysis, we modeled scattering as a Gaussian kernel with an effective width of approximately  \textasciitilde2 $\mu$m. This scattering causes photons emitted from regions just adjacent to the focal position along the z-axis---where the beam is naturally broader---to contribute to the signal at the focus. Consequently, Gaussian fits to the transverse intensity cross‐sections yield a beam waist that is larger than the true value. This effect could account for the observed discrepancy between the expected and measured $w_0$.

Additionally, spherical aberrations induced by the glass cover at the entrance of the water tank may further distort the beam profile and contribute to the apparent broadening.

\subsection{Thermal lens model}

\begin{figure}[H]
    \centering
    \includegraphics[width=1\linewidth]{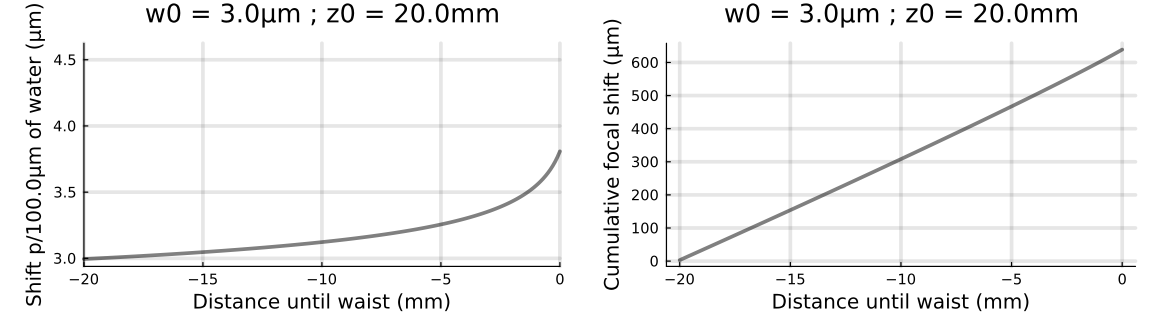}
    \caption{Focal shift induced per 100 $\mu$m-thick water layer. Left: Contribution as a function of the layer’s position relative to the final focal point. Right: Cumulative total focal shift.}
    \label{fig_FocalShiftPerLayer}
\end{figure}

    \label{fig:Optical_setup}
\end{document}